\documentclass[journal,twoside,web]{ieeecolor}
\usepackage{generic}
\usepackage{pifont}
\usepackage{cite}
\usepackage{multirow}
\usepackage{array}
\usepackage{amsmath,amssymb,amsfonts}
\usepackage{algorithmic}
\usepackage{booktabs} 
\usepackage{graphicx}
\usepackage{algorithm,algorithmic}
\usepackage{hyperref}
\usepackage[table]{xcolor}
\usepackage{booktabs}
\hypersetup{hidelinks=true}
\usepackage{textcomp}
\def\BibTeX{{\rm B\kern-.05em{\sc i\kern-.025em b}\kern-.08em
    T\kern-.1667em\lower.7ex\hbox{E}\kern-.125emX}}
\begin{document}
\title{ECG-Expert-QA: A Benchmark for Evaluating Medical Large Language Models in Heart Disease Diagnosis}

\author{Xu Wang, Jiaju Kang, Puyu Han, Yubao Zhao, Qian Liu, \\Liwenfei He, Lingqiong Zhang, Lingyun Dai, Yongcheng Wang, Jie Tao
\thanks{This work was supported by the “Pioneer” R\&D Program of Zhejiang Province under Grant No. 2024C03005. The first institutional affiliation for this study is Zhejiang University School of Medicine, The First Affiliated Hospital \& Liangzhu Laboratory. \textit{Corresponding author: Jie Tao; Yongcheng Wang}.}%
\thanks{Xu Wang is with the Department of Laboratory Medicine of The First Affiliated Hospital \& Liangzhu Laboratory, Zhejiang University School of Medicine, Hangzhou 310000, China; 
the School of Computer Science and Technology, Shandong Jianzhu University, Jinan 250000, China 
and the FUXI AI Lab, Shenzhen 518000, China (e-mail: 202311102025@stu.sdjzu.edu.cn).}%
\thanks{Jiaju Kang is with Beijing Normal University, Zhuhai Campus, Zhuhai 519087, China and also with the FUXI AI Lab, Shenzhen 518000, China (e-mail: kangjiaju@fuxi-lab.com).}%
\thanks{Puyu Han is with Southern University of Science and Technology, Shenzhen 518055, China and also with the FUXI AI Lab, Shenzhen 518000, China (e-mail: 12432627@mail.sustech.edu.cn ).}%
\thanks{Yubao Zhao is with the China University of Geosciences (Wuhan), Wuhan 430000, China (e-mail: 20211000862@cug.edu.cn).}%
\thanks{Qian Liu, Liwenfei He, Lingqiong Zhang, and Lingyun Dai are with Hangzhou Baorun Biotechnology Co., Ltd., Hangzhou, China (e-mail: margaretlq@baorunbiotech.com; heliwenfei@baorunbiotech.com; zhanglingqiong@baorunbiotech.com; kellydai@baorunbiotech.com).}%
\thanks{Yongcheng Wang is also with the Department of Laboratory Medicine of The First Affiliated Hospital \& Liangzhu Laboratory, Zhejiang University School of Medicine, Hangzhou 310000, China (e-mail:  yongcheng@zju.edu.cn).}%
\thanks{Jie Tao is with the Department of Laboratory Medicine of The First Affiliated Hospital \& Liangzhu Laboratory, Zhejiang University School of Medicine, Hangzhou 310000, China (e-mail: taojie1991@zju.edu.cn).}%
}
 
\maketitle

\begin{abstract}
We present ECG-Expert-QA, a comprehensive multimodal dataset for evaluating diagnostic capabilities in electrocardiogram (ECG) interpretation. It combines real-world clinical ECG data with systematically generated synthetic cases, covering 12 essential diagnostic tasks and totaling 47,211 expert-validated QA pairs. These encompass diverse clinical scenarios, from basic rhythm recognition to complex diagnoses involving rare conditions and temporal changes.
A key innovation is the support for multi-turn dialogues, enabling the development of conversational medical AI systems that emulate clinician-patient or interprofessional interactions. This allows for more realistic assessment of AI models' clinical reasoning, diagnostic accuracy, and knowledge integration.
Constructed through a knowledge-guided framework with strict quality control, ECG-Expert-QA ensures linguistic and clinical consistency, making it a high-quality resource for advancing AI-assisted ECG interpretation. It challenges models with tasks like identifying subtle ischemic changes and interpreting complex arrhythmias in context-rich scenarios.
To promote research transparency and collaboration, the dataset, accompanying code, and prompts are publicly released at \underline{https://github.com/Zaozzz/ECG-Expert-QA}.
% Our dataset is open-source and available at \underline{https://github.com/Zaozzz/ECG-Expert-QA}.
\end{abstract}

\begin{IEEEkeywords}
ECG Interpretation, Medical Large Language Models, Multi-turn QA, Cross-modal Clinical Reasoning, Ethical and Risk-aware Evaluation
% Enter key words or phrases in alphabetical order, separated by commas. Using the IEEE Thesaurus can help you find the best standardized keywords to fit your article. Use the thesaurus access request form for free access to the IEEE Thesaurus: \underline{https://www.ieee.org/publications/services/thesaurus-acce}\\
% \underline{ss-page.com.}
\end{IEEEkeywords}

\section{Introduction}
\label{sec1}

\IEEEPARstart{E}{CG} interpretation plays a vital role in cardiovascular diagnosis and clinical decision-making. However, accurate interpretation often relies on expert-level reasoning and domain-specific knowledge, which poses challenges in scenarios with limited medical resources. As large language models (LLMs) become increasingly powerful, there is growing interest in their potential to augment or automate medical tasks. Yet, the application of these models to ECG analysis remains underdeveloped, hindered by limitations in evaluation frameworks, data diversity, and ethical oversight.

\begin{figure}[h]
    \centering
    \includegraphics[width=1\linewidth]{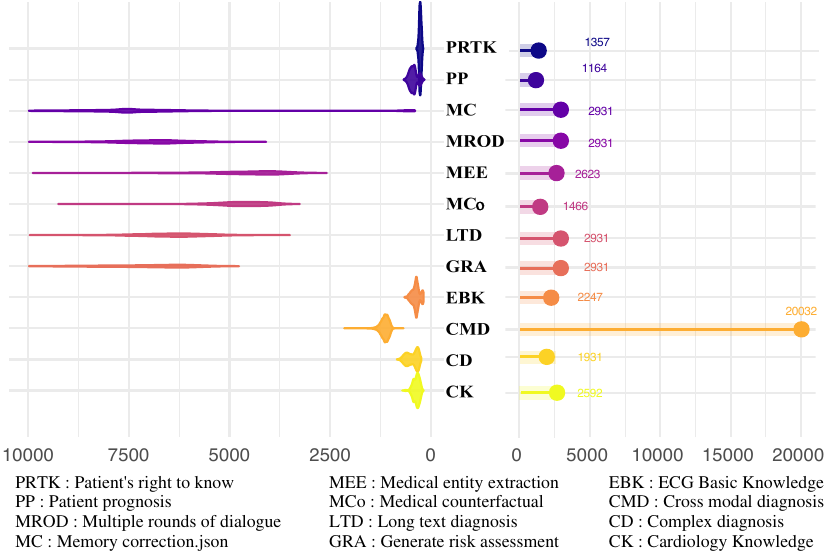}
    \label{fig1}
    \caption{Task Distribution and Sample Volume in ECG-Expert-QA Benchmark}
\end{figure}

\begin{table*}[h]
\centering
\caption{Comparison of ECG-related Benchmarks in Terms of Dialogue, Reasoning, and Patient-centric Dimensions}
\renewcommand{\arraystretch}{1.2}
\begin{tabular}{lccccc}
\toprule
\textbf{Dataset / System} & \textbf{Dialogue Support} & \textbf{Clinical Reasoning} & \textbf{Rare Disease} & \textbf{Multi-turn} & \textbf{Prognosis / Informed} \\
\midrule
ECG-Chat & Single-turn & Partial & Limited & \ding{56} & No \\
ECG-QA & Single-turn & Template-based & No & \ding{56} & No \\
GEM & -- & Contextual modeling & Unknown & \ding{56} & Temporal ECG modeling \\
MedGemini & -- & Advanced reasoning & Yes & \ding{56} & Partial prognosis \\
MedQA-CS & Multi-turn & Structured QA & Yes & \ding{52} & No \\
AgentClinic & Multi-turn & Interactive logic & Simulated & \ding{52} & Simulated patient awareness \\
PTB-XL & -- & Classification only & Limited & \ding{56} & No \\
\textbf{OURS} & Multi-turn & Full reasoning & Yes & \ding{52} & Prognosis \& doctor-patient dialogue \\
\bottomrule
\end{tabular}
\label{tab1}
\end{table*}

Despite advances in explainable AI, most current approaches fail to support high-stakes diagnostic reasoning. Holzinger et al. have emphasized that for AI to be trusted in clinical practice, transparency and interpretability must be designed from the outset \cite{holzinger2017needbuildexplainableai}. At the same time, large-scale datasets such as MIMIC-IV have laid the foundation for training medical AI models with real-world clinical data \cite{johnson2023mimiciv}. However, these datasets lack structured evaluation protocols for measuring reasoning, ethical decision-making, or interactional capabilities within ECG contexts.

Recent work has shown that LLMs can help bridge care gaps in underserved regions. Strika et al. illustrated how AI and LLMs might mitigate disparities in “medical deserts,” while also cautioning against linguistic bias and ethical blind spots in current systems \cite{strika2024bridging}. Domain-specific models such as BioGPT have shown promise in biomedical text generation \cite{Luo_2022}, yet remain underexplored in diagnostic scenarios involving multimodal and temporal reasoning. Ethical concerns also persist, as noted by Morley et al., who stress the importance of building AI systems that can operate responsibly across levels of abstraction—from individual to institutional decision-making \cite{MORLEY2020113172}. Moreover, efforts to align LLMs with clinical data, as seen in Yang et al.'s work on EHR-trained models, suggest that more tailored evaluation datasets are needed to ensure performance, fairness, and safety \cite{yang2022llmEHR}.

To address these gaps, we present ECG-Expert-QA, a novel benchmark designed to evaluate the diagnostic reasoning, ethical judgment, and language understanding capabilities of medical LLMs in the context of ECG interpretation. Unlike prior datasets, ECG-Expert-QA incorporates diverse, high-complexity cases, multi-turn dialogues, and risk-sensitive scenarios to support comprehensive and scalable evaluation. This work aims to enable the development of more trustworthy and context-aware LLMs for real-world medical applications.

\section{Related Work}
\label{sec2}
\begin{figure*}[h]
    \centering
    \includegraphics[width=1\linewidth]{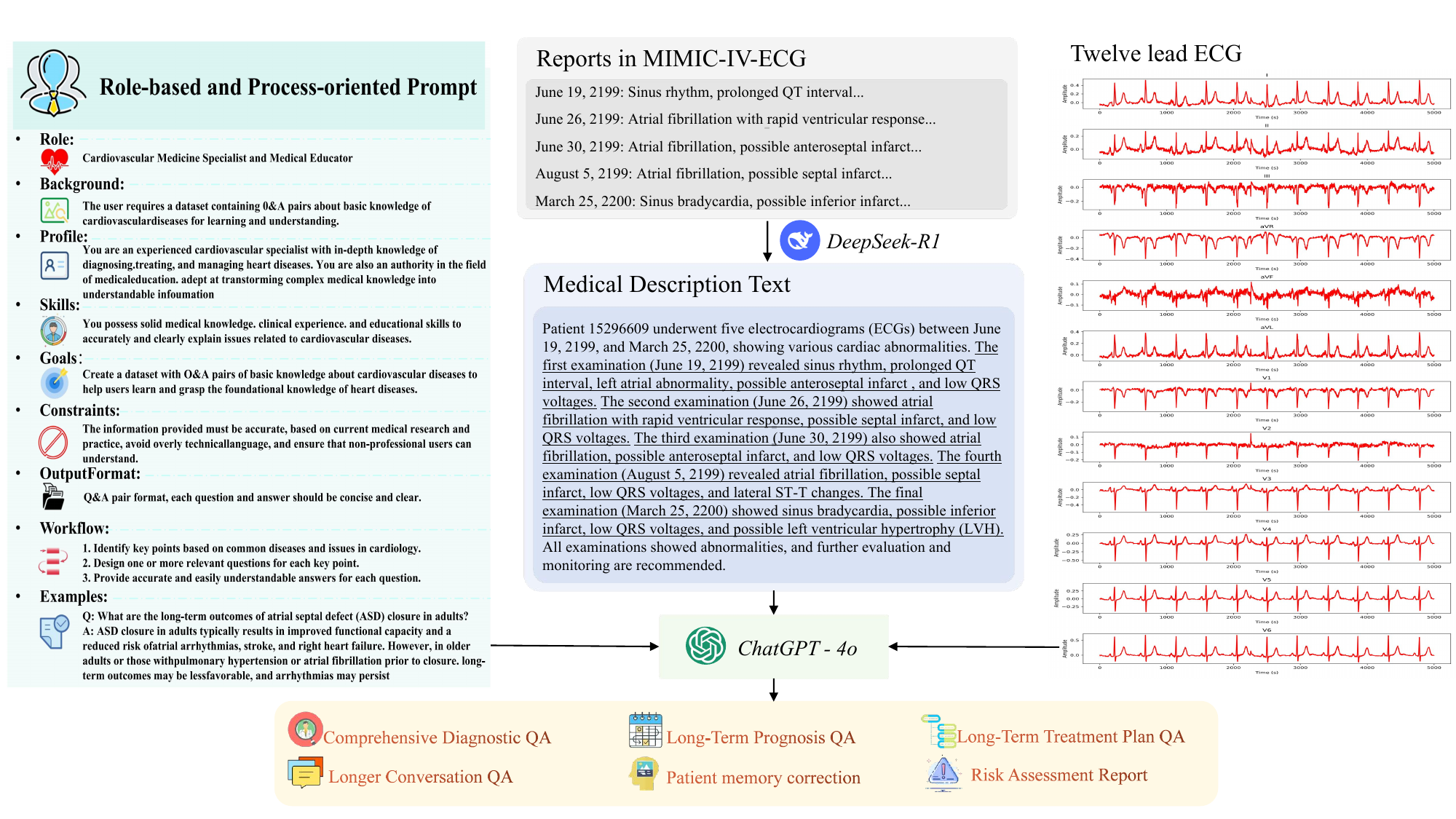}
    \label{fig2}
    \caption{A Role-Based and Process-Oriented Framework for ECG Multimodal QA Generation}
\end{figure*}

With the rapid advancement of LLMs and multimodal learning, automated ECG interpretation has progressed beyond conventional classification tasks to encompass clinical reasoning and dialog-based medical QA. A number of high-quality multimodal datasets have emerged, aiming to bridge the gap between AI and real-world clinical decision-making, especially in the context of ECG-based diagnosis.

One of the earlier efforts, ECG-QA~\cite{bib1}, proposed a question-answering framework for ECG interpretation. However, its structure is limited to single-turn QA with fixed question templates, lacking the capacity to incorporate contextual or longitudinal clinical reasoning. Later, ECG-Chat~\cite{bib2} introduced a large-scale ECG-language model trained on 19k diagnostic ECG samples, demonstrating the promise of integrating signal data with natural language. Nonetheless, it remains constrained in handling conversational dynamics and rare disease scenarios.

To enhance generalizability and clinical applicability, several studies have explored fusing ECG signals with text, images, and structured data within multimodal frameworks. For instance, GEM~\cite{bib3} proposed a multimodal diagnostic model that combines visual and temporal ECG information for more robust and explainable reasoning. Similarly, MedGemini~\cite{bib4} offered a family of multimodal medical foundation models capable of processing ECG, imaging, and structured EHRs for comprehensive clinical understanding.

Recent efforts have also extended ECG modeling from classification to natural language report generation and QA. ECG-ReGen~\cite{bibReGen} presented a retrieval-augmented framework that integrates self-supervised ECG representation learning with large language models to generate clinical reports and perform diagnostic question answering. While promising, it primarily supports single-turn interactions and lacks support for conversational reasoning or patient-specific context.

In parallel, benchmarking efforts based on large-scale ECG datasets have also advanced. One notable example is the PTB-XL~\cite{bibPTBXL} dataset, which provides a publicly available, richly annotated 12-lead ECG corpus for various diagnostic tasks. Accompanied by rigorous benchmarking and analysis across ResNet- and Inception-based deep learning models, PTB-XL has contributed significantly to structured performance evaluation in ECG classification. However, while valuable for model training and general benchmarking, such datasets remain focused on static classification tasks and lack support for conversational modeling or contextual reasoning.

Moreover, the diagnostic value of ECG-based QA systems is inherently tied to signal quality. A comprehensive review~\cite{bibQualityReview} summarized recent advances in automated ECG quality assessment, underscoring the importance of reliable signal input for downstream AI-driven analysis. Incorporating signal quality metrics into future conversational benchmarks remains a valuable research direction.

Concurrently, the need to evaluate medical LLMs in realistic diagnostic settings has led to the development of interactive and multi-agent environments such as AgentClinic~\cite{bib5} and MedQA-CS~\cite{bib6}. While these benchmarks enable evaluation across diverse clinical domains, they are often focused on textual data or radiology tasks, leaving ECG interpretation underrepresented in the multimodal conversational AI landscape.

Additionally, datasets like Human3.6M~\cite{bibH36M1,bibH36M2} (pose-text), AudioSet~\cite{bibAudioSet} (audio-text), and LuoJiaHOG~\cite{bibLuoJiaHOG} (image-text) have demonstrated success in other domains, furthering the potential of multimodal learning. These datasets pair different modalities, including motion, sound, and visual data, with textual descriptions, providing valuable insights into how multimodal models can integrate diverse data sources for improved performance.

ECG-Expert-QA addresses these limitations by introducing the first open-source benchmark to support multi-turn ECG diagnostic QA. By integrating real clinical records with systematically synthesized cases across 12 diagnostic tasks and 47,211 QA pairs, it supports rich clinical contexts, rare conditions, and temporal disease progression patterns. Its design enables comprehensive evaluation of medical LLMs in terms of diagnostic accuracy, clinical reasoning, and knowledge integration, setting a new standard for conversational ECG AI systems.

\section{Methods}
\label{sec3}

This study introduces three primary dataset generation strategies: expert knowledge-guided professional knowledge assessment, cross-modal diagnosis in complex medical environments, and medical risk assessment. Each strategy is designed to address specific challenges in intelligent ECG interpretation by generating high-quality question-answer (QA) pairs. The expert-guided approach leverages domain-specific medical expertise to steer LLMs in generating accurate and clinically relevant content. The cross-modal diagnosis approach converts semi-structured clinical data, such as ECG signals, into detailed textual narratives to support reasoning-based QA generation. The medical risk assessment approach focuses on ethical considerations, counterfactual scenarios, and patient rights, producing QA pairs that simulate realistic, high-stakes clinical situations. Each method employs distinct input modalities and interaction protocols to facilitate medical knowledge learning, pathological reasoning, and multi-turn dialogue, thereby enhancing the robustness and applicability of LLMs in real-world clinical settings.

\begin{figure*}[h]
    \centering
    \includegraphics[width=0.84\linewidth]{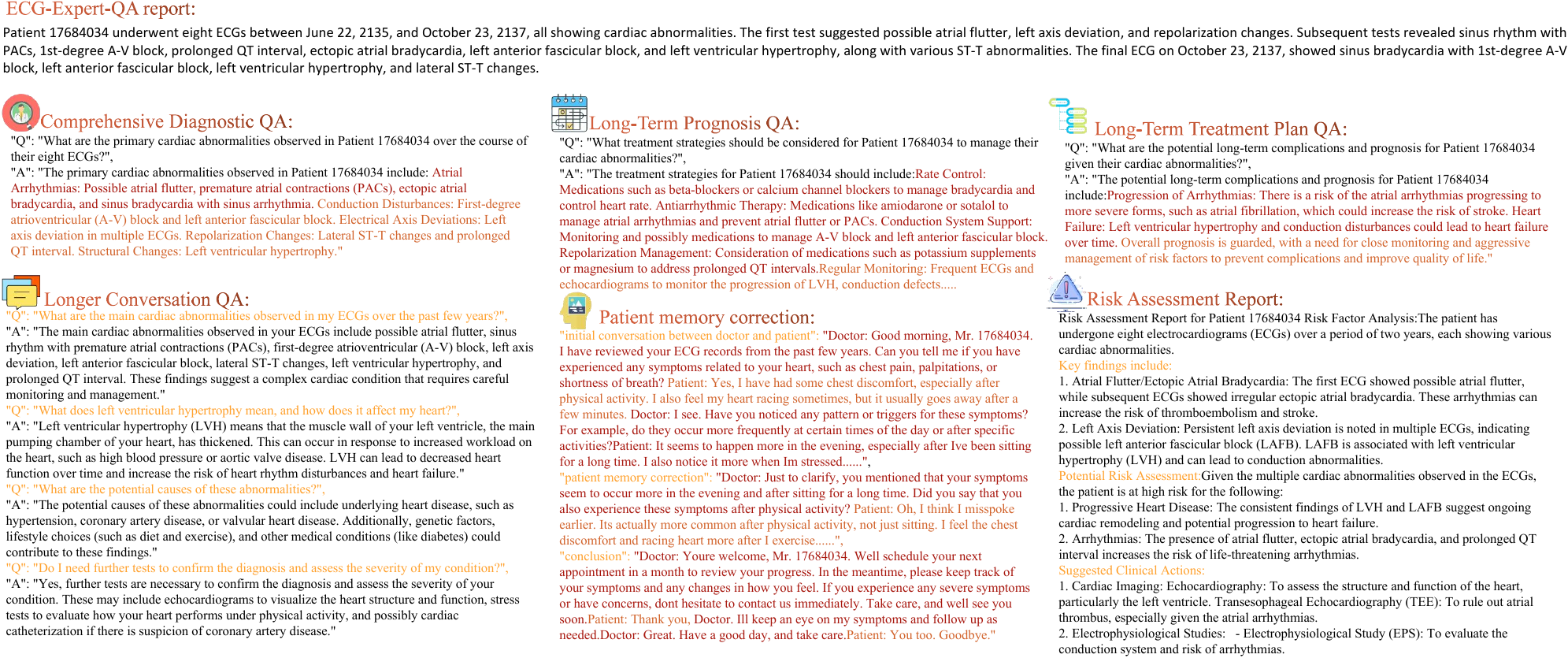}
    \label{fig3}
    \caption{ECG-Expert-QA Multi-Task QA Examples Across Cardiac Cases - 1}
\end{figure*}

\begin{figure*}[h]
    \centering
    \includegraphics[width=0.84\linewidth]{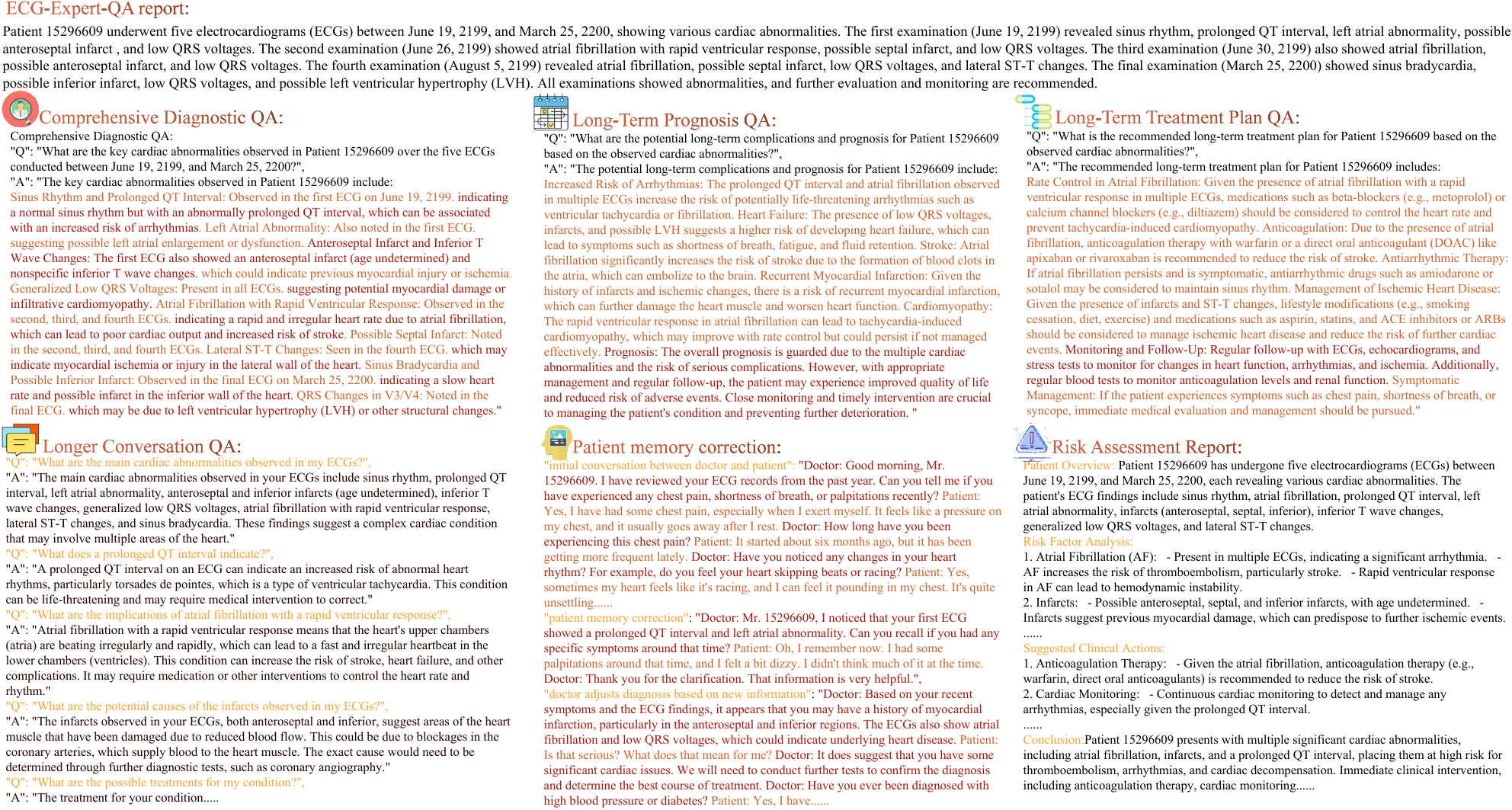}
    \label{fig4}
    \caption{ECG-Expert-QA Multi-Task QA Examples Across Cardiac Cases - 2}
\end{figure*}

\begin{figure*}[h]
    \centering
    \includegraphics[width=0.84\linewidth]{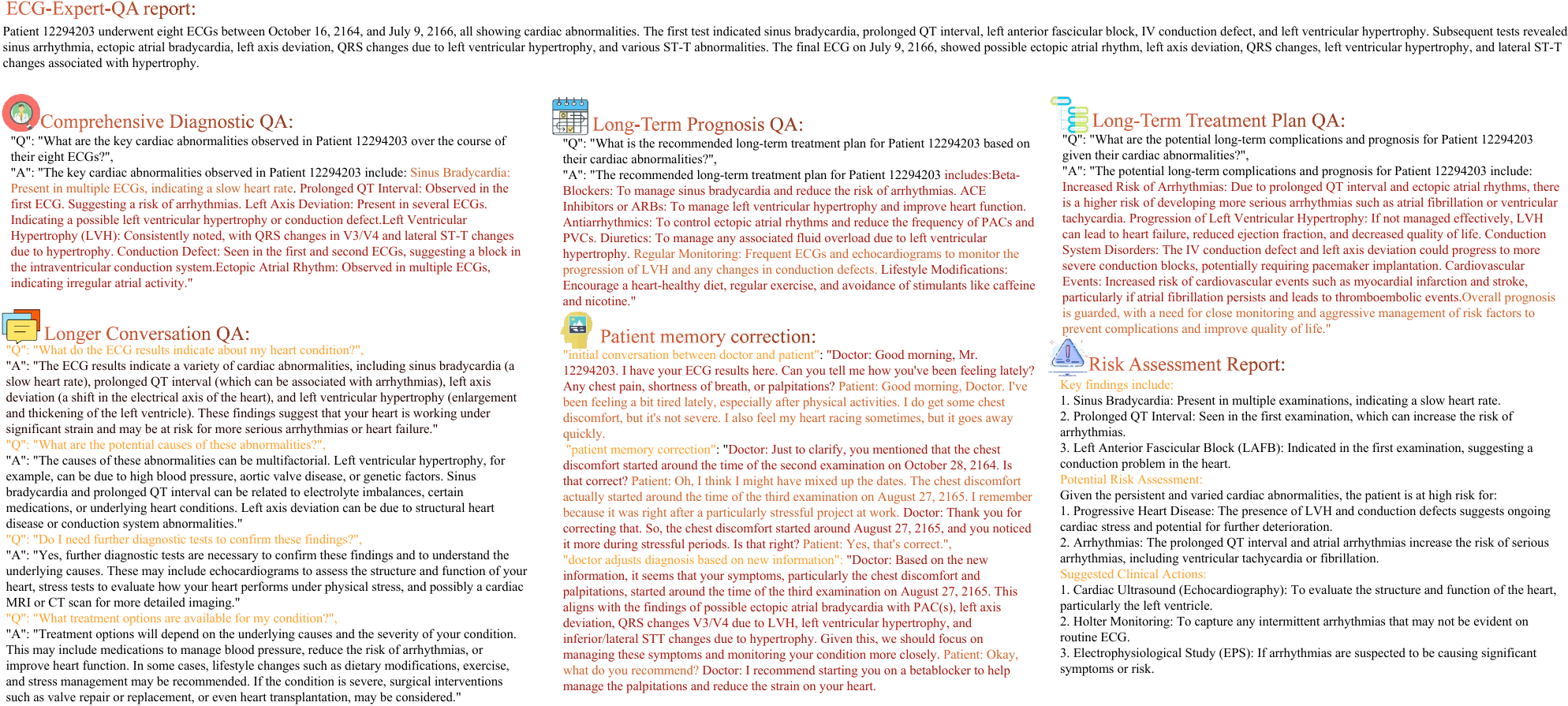}
    \label{fig5}
    \caption{ECG-Expert-QA Multi-Task QA Examples Across Cardiac Cases - 3}
\end{figure*}

\begin{figure*}[h]
    \centering
    \includegraphics[width=0.88\linewidth]{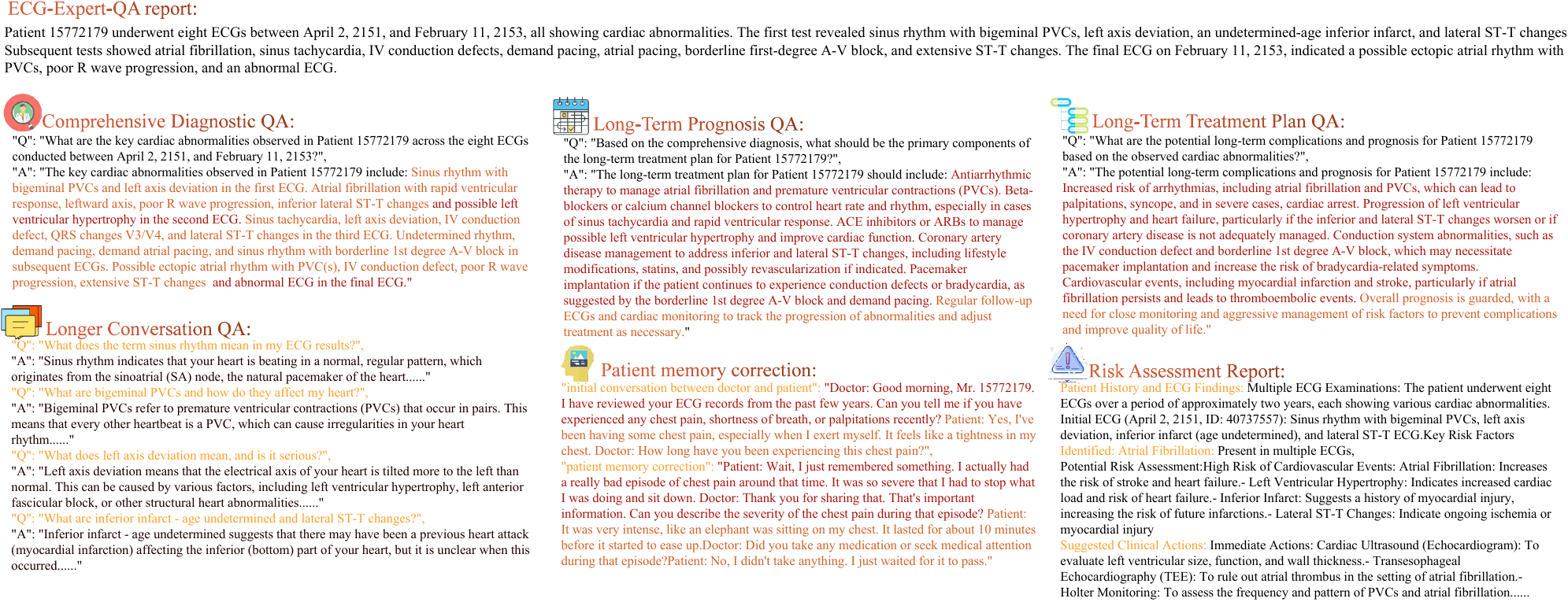}
    \label{fig6}
    \caption{ECG-Expert-QA Multi-Task QA Examples Across Cardiac Cases - 4}
\end{figure*}

\begin{figure*}[h]
    \centering
    \includegraphics[width=0.88\linewidth]{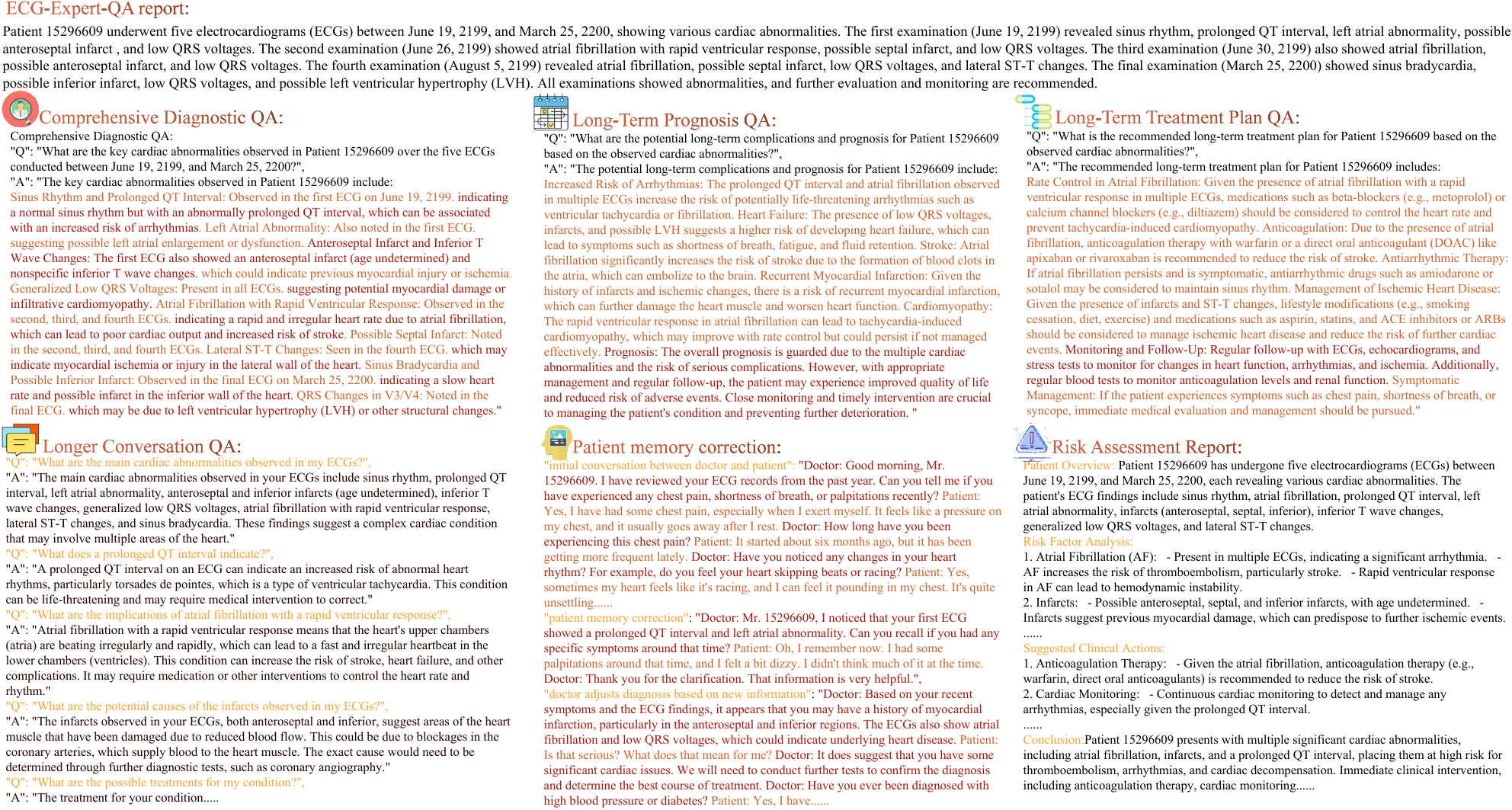}
    \label{fig7}
    \caption{ECG-Expert-QA Multi-Task QA Examples Across Cardiac Cases - 5}
\end{figure*}

\begin{figure*}[h]
    \centering
    \includegraphics[width=0.31\linewidth]{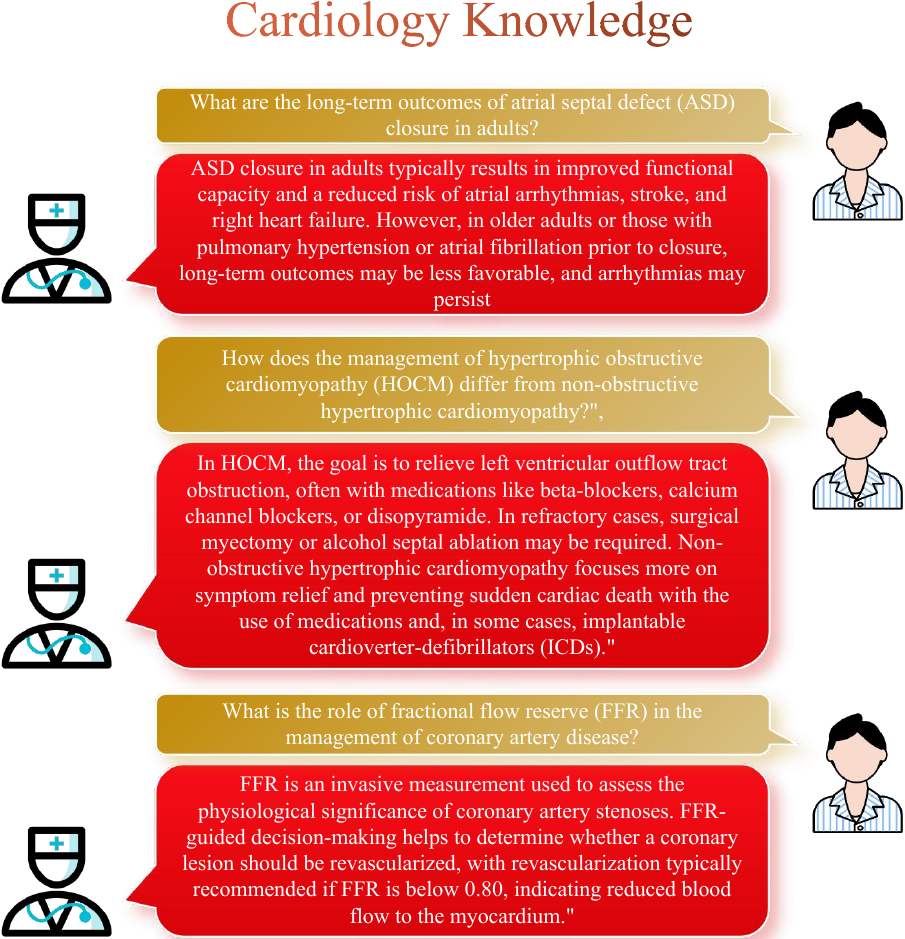}
    \includegraphics[width=0.27\linewidth]{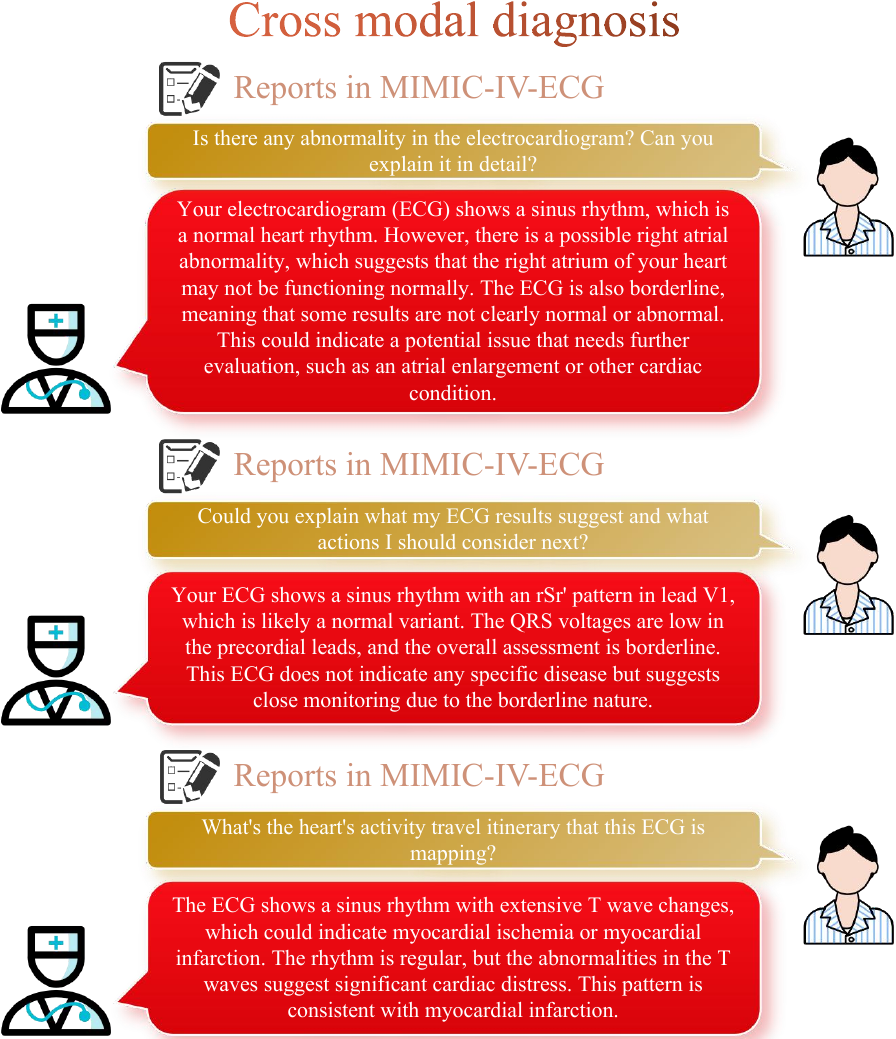}
    \includegraphics[width=0.31\linewidth]{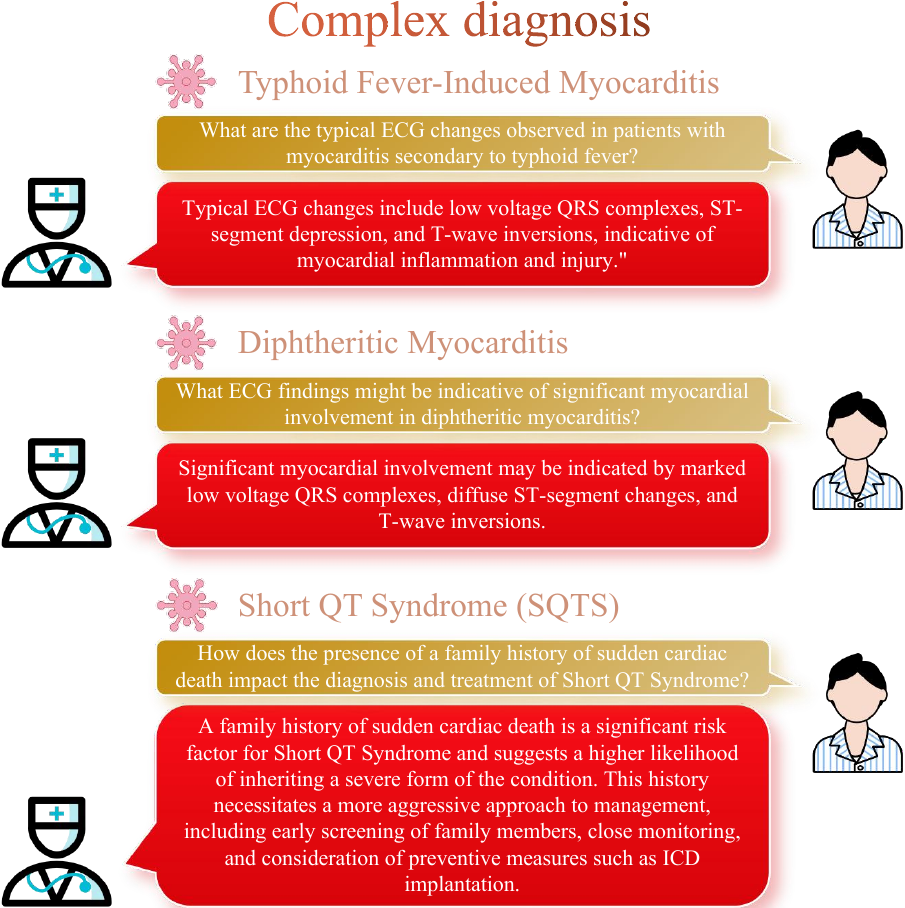}
    \label{fig8}
    \caption{ECG-Expert-QA Multi-Task QA Examples Across Cardiac Cases - 6}
\end{figure*}

\subsection*{Expert Knowledge-Guided Professional Knowledge Assessment}

This method constructs a high-quality medical QA corpus through continuous interaction with state-of-the-art LLMs (e.g., GPT-4o\cite{bib13}). Local deployment or API-based access is used to integrate domain expertise via role-based prompting strategies that simulate realistic doctor–patient dialogues. The resulting data spans the following core areas:

\begin{itemize}
\item \textbf{CK (Cardiology Knowledge)}: Covers foundational cardiology concepts, including cardiac anatomy, function, common cardiovascular diseases (e.g., coronary artery disease, heart failure), etiology, symptoms, diagnostics, and treatment pathways. It is generated by designing multi-turn dialogue prompts that guide LLMs to produce diverse responses for knowledge distillation.

\item \textbf{EBK (ECG Basic Knowledge)}: Focuses on electrocardiographic principles, waveform interpretation, detection of common abnormalities such as premature ventricular contractions and atrial fibrillation, and practical ECG usage in clinical contexts. This module is also created through iterative multi-turn dialogue with prompt variations to obtain refined and diverse answers.

\item \textbf{CD (Complex Diagnosis)}: Involves multi-turn expert discussions simulating complex cardiac scenarios, including comorbidities and rare conditions, aiming to generate in-depth clinical reasoning and individualized treatment strategies. It is constructed by filtering heart-related cases from the GenMedicalEval\cite{bib10,bib11,bib12} dataset and prompting the model to perform diagnostic reasoning.
\end{itemize}

\subsection*{Cross-Modal Diagnosis in Complex Medical Environments}

This method transforms semi-structured ECG data—such as 12-lead ECG signals—into descriptive textual narratives that detail waveform features, abnormal signals, clinical symptoms, and diagnostic insights. These descriptions are then paired with role-specific prompts to guide LLMs in producing multi-faceted QA pairs. Applications include:

\begin{itemize}
\item \textbf{CMD (Cross Modal Diagnosis)}: Integrates ECG imagery and textual data to establish associations between waveform abnormalities and clinical diagnoses, covering pathologies such as arrhythmias and myocardial infarction. It is generated using diagnostic information and reports from the MIMIC-IV-ECG\cite{bib0} dataset, reformulated into structured narratives and paired with carefully designed prompts.

\item \textbf{PP (Patient Prognosis)}: Produces prognosis-based QA pairs by analyzing current cardiac status and forecasting disease trajectories or recovery timelines. This is generated by summarizing ECG-related diagnostic content from MIMIC-IV-ECG and prompting the model to predict outcomes.

\item \textbf{LTD (Long Text Diagnosis)}: Uses extended clinical narratives to evaluate the model’s comprehension and diagnostic ability in handling rich contextual information with multiple variables. These are based on long-form reports from MIMIC-IV-ECG, transformed into textual inputs for reasoning-focused QA generation.

\item \textbf{MROD (Multiple Rounds of Dialogue)}: Simulates ongoing multi-turn doctor–patient interactions, assessing the model’s capacity for maintaining coherent reasoning across conversational contexts. It is built from MIMIC-IV-ECG case data and structured as dialogue scripts to guide multi-turn interactions.

\item \textbf{MC (Memory Correction)}: Evaluates the model’s adaptability by testing its ability to revise or refine previous responses when new medical information is introduced mid-dialogue. This is achieved by incrementally revealing information from MIMIC-IV-ECG cases and observing the model’s ability to adjust its responses.

\item \textbf{MEE (Medical Entity Extraction)}: Applies natural language processing techniques to extract structured medical entities (e.g., symptoms, diagnoses, treatments) from unstructured ECG-related text. These QA pairs are generated by inputting ECG-based clinical reports from MIMIC-IV-ECG and prompting the model to extract key medical elements.
\end{itemize}

\subsection*{Medical Risk Assessment}

This module addresses ethical and legal dimensions of medical practice, including patient autonomy, diagnostic uncertainty, and hypothetical scenario analysis. Role-based interaction frameworks are applied to simulate ethically charged conversations using LLMs. QA pairs are generated across the following categories:

\begin{itemize}
\item \textbf{PRTK (Patient's Right to Know)}: Simulates scenarios focused on informed consent, shared decision-making, and transparent communication, highlighting the legal and ethical obligations of clinicians. This is generated through multi-turn dialogue prompting that explores ethically nuanced doctor–patient interactions.

\item \textbf{MCo (Medical Counterfactual)}: Constructs counterfactual diagnostic cases by altering variables such as treatment choices or drug dosages, then analyzing the impact on patient outcomes. It is generated by modifying variables in MIMIC-IV-ECG cases and prompting the model to reason through the consequences.

\item \textbf{GRA (Generate Risk Assessment)}: Generates clinical risk assessments that synthesize multiple data points to evaluate the potential risks and benefits of proposed interventions. This is constructed using rich diagnostic reports from MIMIC-IV-ECG and prompts that elicit comprehensive risk-benefit evaluations.
\end{itemize}

\subsection*{Summary}

Through the application of three systematic dataset generation approaches—expert-guided knowledge assessment (CK, EBK, CD), cross-modal diagnostic reasoning (CMD, PP, LTD, MROD, MC, MEE), and medical risk evaluation (PRTK, MCo, GRA)—this study constructs a comprehensive and high-quality dataset for intelligent ECG interpretation. These datasets support not only factual knowledge acquisition and diagnostic reasoning, but also ethical decision-making and realistic conversational modeling. By enhancing the adaptability and accuracy of large language models in dynamic clinical environments, the proposed framework lays a strong foundation for the development of trustworthy, multimodal, and patient-centered medical AI systems.

\section{Experiments}
\label{sec4}

To evaluate the performance of LLMs on the ECG-Expert-QA dataset, we adopted four widely used evaluation metrics in natural language generation (NLG): \textbf{BLEU-1}\cite{bib7}, \textbf{ROUGE-L}\cite{bib8}, and \textbf{METEOR}\cite{bib9}. These metrics measure the similarity between the model-generated answers and ground-truth references in terms of lexical overlap and semantic relevance.

\subsection*{BLEU (Bilingual Evaluation Understudy)}

BLEU is a precision-based metric that measures the proportion of overlapping n-grams between the generated output and the reference answer. In our evaluation, we focus on **BLEU-1**, which uses unigram (single word) precision to reflect basic word-level alignment and lexical accuracy.

\begin{equation}
\text{BLEU-1} = \text{BP} \cdot \exp\left( \log p_1 \right) = \text{BP} \cdot p_1
\end{equation}

where:
\begin{itemize}
\item $p_1$ is the modified unigram precision,
\item BP is the brevity penalty:
\end{itemize}

\begin{equation}
\text{BP} =
\begin{cases}
1 & \text{if } c > r \\
e^{(1 - \frac{r}{c})} & \text{if } c \le r
\end{cases}
\end{equation}

Here, $c$ is the length of the candidate sentence and $r$ is the length of the reference sentence. The brevity penalty ensures that excessively short outputs are penalized, promoting more complete responses.

\begin{figure*}[h]
    \centering
    \includegraphics[width=1\linewidth]{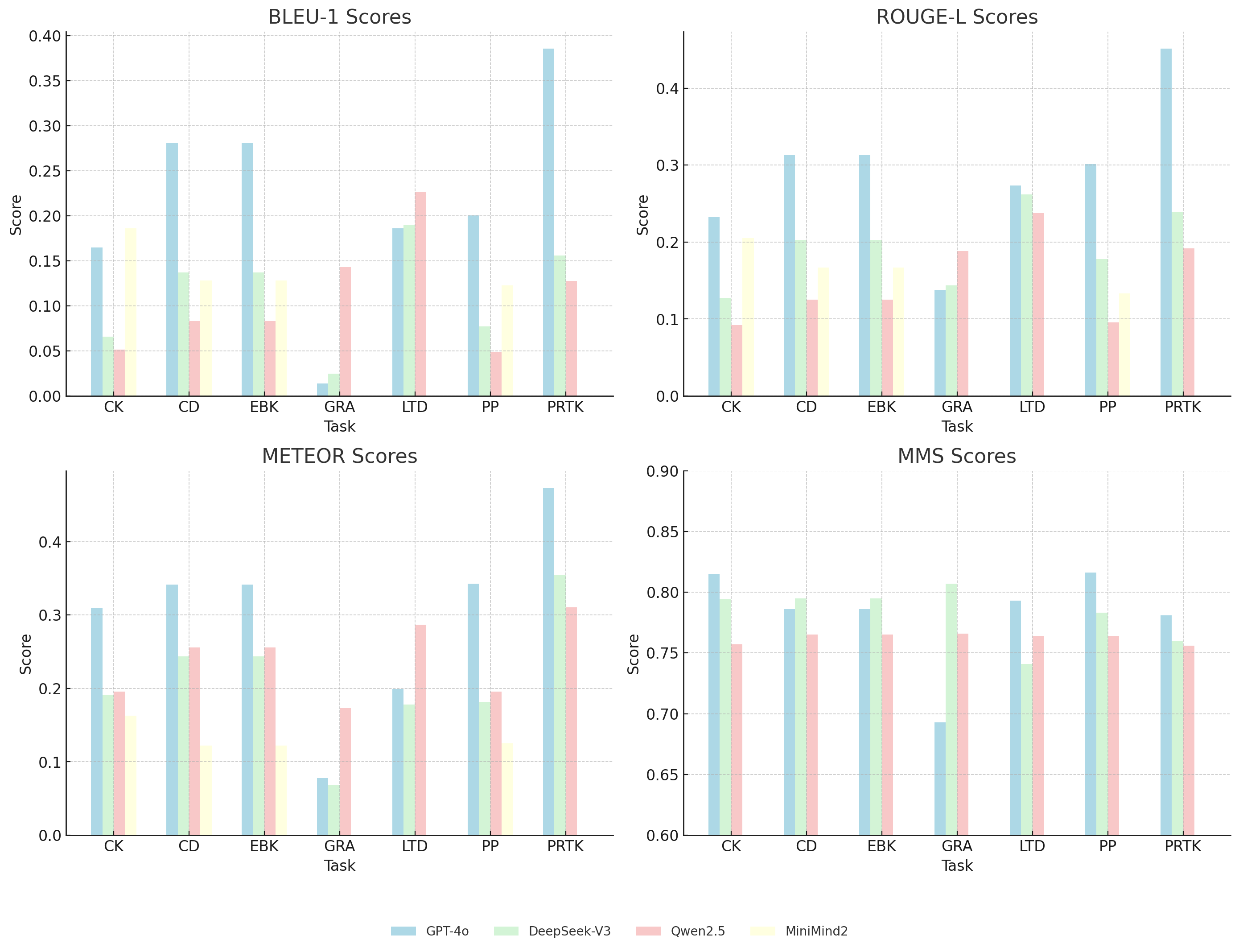}
    \label{fig9}
    \caption{Comparative Performance of LLMs on ECG-Expert-QA. Color Mapping: Blue – GPT-4o, Orange – DeepSeek-V3,  Green – Qwen2.5}
\end{figure*}

\subsection*{ROUGE-L (Recall-Oriented Understudy for Gisting Evaluation)}

ROUGE-L is a recall-oriented metric based on the longest common subsequence (LCS) between the generated text and the reference. It considers both precision and recall, and is particularly useful for evaluating coverage of reference content.

\begin{equation}
\text{ROUGE-L} = F_\beta = \frac{(1 + \beta^2) \cdot \text{Precision} \cdot \text{Recall}}{\text{Recall} + \beta^2 \cdot \text{Precision}}
\end{equation}

where:
\begin{itemize}
\item $\text{Precision} = \frac{LCS}{\text{length of candidate}}$
\item $\text{Recall} = \frac{LCS}{\text{length of reference}}$
\item $\beta$ is typically set to 1.
\end{itemize}

\subsection*{METEOR (Metric for Evaluation of Translation with Explicit ORdering)}

METEOR incorporates both precision and recall while also considering synonym matching, stemming, and paraphrasing, making it more semantically aware than BLEU or ROUGE.

\begin{equation}
\text{METEOR} = F_{\text{mean}} \cdot (1 - \text{Penalty})
\end{equation}

where:
\begin{equation}
F_{\text{mean}} = \frac{10 \cdot P \cdot R}{9P + R}
\end{equation}

\begin{equation}
\text{Penalty} = 0.5 \cdot \left( \frac{\# \text{chunks}}{\# \text{matches}} \right)^3
\end{equation}

Here, $P$ and $R$ are unigram precision and recall, respectively, and ``chunks'' refers to the number of contiguous matched word sequences.

\subsection*{Task-Specific Evaluation}

The above metrics were applied to seven sub-datasets extracted from ECG-Expert-QA, each representing a clinically meaningful task: Cardiology Knowledge, Complex Diagnosis, ECG Basic Knowledge, Generate Risk Assessment, Long Text Diagnosis, Patient Prognosis and Patient’s Right to Know.

\subsection*{Model-to-Model Scoring}

In addition to these standard metrics, we also introduced a Model-to-Model Scoring (MMS) mechanism. This method involves using one large model (Model A) to evaluate the performance of another model (Model B) based on its generated answers and the ground-truth reference\cite{bibmms1,bibmms2}. Specifically, Model A compares the response from Model B with the correct answer, and assigns a score based on factors such as semantic consistency, content coverage, and language fluency.

MMS offers an automated way to evaluate model outputs, reducing the need for human intervention and helping to optimize Model B’s performance. By leveraging the strengths of one model to assess the performance of another, we gain a deeper understanding of each model’s strengths and weaknesses across different tasks.

\subsection*{Lightweight Model Fine-tuning}

To further evaluate the applicability of our benchmark to lightweight models, we fine-tune MiniMind2\cite{bibminimind}, a compact LLM with only 25.8 million parameters, on the ECG-Expert-QA dataset. Although MiniMind2 contains significantly fewer parameters than larger-scale models, our goal is to assess whether ECG-Expert-QA can still enable meaningful diagnostic performance under constrained model capacity.

Due to input/output token limitations inherent to MiniMind2, we restrict its evaluation to selected sub-datasets that feature relatively short contexts and responses. This design allows us to explore the feasibility of deploying lightweight medical LLMs in real-world, resource-constrained scenarios, such as mobile or embedded systems.

\subsection*{Summary}

Through a combination of standard NLG metrics and model-to-model evaluation, we comprehensively assessed the diagnostic and linguistic capabilities of multiple LLMs on ECG-Expert-QA. In addition to benchmarking large models, we also evaluated the lightweight MiniMind2 to examine the dataset’s adaptability under constrained model capacity. The results demonstrate that ECG-Expert-QA supports robust performance evaluation across a wide spectrum of model sizes and provides valuable insights into the trade-offs between model complexity and clinical applicability.

\section{Result}
\label{sec5}
In this section, we present the evaluation results of four models—GPT-4o, DeepSeek-V3, Qwen2.5, and MiniMind2—on seven clinically relevant tasks from the ECG-Expert-QA dataset. The models were assessed using three widely adopted natural language generation metrics: \textbf{BLEU-1}, \textbf{ROUGE-L}, and \textbf{METEOR}, as well as the newly introduced \textbf{Model-to-Model Scoring (MMS)}. These metrics allow for a comprehensive evaluation of both lexical overlap and semantic alignment with expert-annotated references.

While GPT-4o, DeepSeek-V3, and Qwen2.5 were evaluated across all tasks, MiniMind2—a lightweight model with only 25.8M parameters—was assessed on a subset of tasks (CK, CD, EBK, PP) due to input/output limitations. The inclusion of MiniMind2 enables exploration of model performance in resource-constrained scenarios.

Through this multi-dimensional evaluation, we aim to assess the models' ability to generate accurate, fluent, and clinically relevant responses. The results highlight each model’s strengths and limitations in terms of diagnostic reasoning, contextual understanding, and communicative precision, offering insight into their applicability in real-world medical AI systems.

% ----- Table for Task CK -----
\begin{table}[h]
\centering
\caption{Evaluation Metrics for Task CK}
\label{tab:metrics_ck}
\begin{tabular}{lcccc}
\toprule
Model & BLEU-1 & ROUGE-L & METEOR & MMS \\
\midrule
GPT-4o        & 0.1648 & \cellcolor[rgb]{0.7,1,0.7}0.2323 & \cellcolor[rgb]{0.7,1,0.7}0.3098 & \cellcolor[rgb]{0.7,1,0.7}0.815 \\
DeepSeek-V3   & 0.0661 & 0.1277 & 0.1917 & 0.794 \\
Qwen2.5       & 0.0514 & 0.0922 & 0.1957 & 0.757 \\
MiniMind2     & \cellcolor[rgb]{0.7,1,0.7}0.1864 & 0.2051 & 0.1631 & --    \\
\bottomrule
\end{tabular}
\end{table}

In the CK task, GPT-4o outperforms all other models across BLEU-1, ROUGE-L, METEOR, and MMS, achieving an MMS of 0.815, reflecting strong alignment with expert references in both lexical and semantic dimensions. DeepSeek-V3 and Qwen2.5 show moderate performance, particularly underperforming in BLEU-1. MiniMind2, however, stands out with the highest BLEU-1 score (0.1864), indicating strong word-level overlap but lower ROUGE-L and METEOR scores, suggesting weaker structural and semantic depth. Its high lexical precision hints at usefulness in lightweight or constrained settings. Overall, GPT-4o remains the most balanced and reliable model for medical text generation, while MiniMind2 excels in concise lexical matching.

% ----- Table for Task CD -----
\begin{table}[h]
\centering
\caption{Evaluation Metrics for Task CD}
\label{tab:metrics_cd}
\begin{tabular}{lcccc}
\toprule
Model & BLEU-1 & ROUGE-L & METEOR & MMS \\
\midrule
GPT-4o        & \cellcolor[rgb]{0.7,1,0.7}0.2810 & \cellcolor[rgb]{0.7,1,0.7}0.3130 & \cellcolor[rgb]{0.7,1,0.7}0.3416 & 0.786 \\
DeepSeek-V3   & 0.1374 & 0.2027 & 0.2435 & \cellcolor[rgb]{0.7,1,0.7}0.795 \\
Qwen2.5       & 0.0833 & 0.1253 & 0.2556 & 0.765 \\
MiniMind2     & 0.1282 & 0.1670 & 0.1219 & --    \\
\bottomrule
\end{tabular}
\end{table}

In the CD task, GPT-4o demonstrates the best overall performance, achieving the highest scores across all metrics, especially in METEOR (0.3416), indicating strong semantic understanding. DeepSeek-V3 follows with moderate results, while Qwen2.5 lags behind. MiniMind2 shows decent BLEU-1 performance (0.1282), suggesting it can capture surface-level information reasonably well, though its lower METEOR score implies limitations in deeper semantic reasoning.

% ----- Table for Task EBK -----
\begin{table}[h]
\centering
\caption{Evaluation Metrics for Task CD}
\label{tab:metrics_cd}
\begin{tabular}{lcccc}
\toprule
Model & BLEU-1 & ROUGE-L & METEOR & MMS \\
\midrule
GPT-4o        & \cellcolor[rgb]{0.7,1,0.7}0.2810 & \cellcolor[rgb]{0.7,1,0.7}0.3130 & \cellcolor[rgb]{0.7,1,0.7}0.3416 & 0.786 \\
DeepSeek-V3   & 0.1374 & 0.2027 & 0.2435 & \cellcolor[rgb]{0.7,1,0.7}0.795 \\
Qwen2.5       & 0.0833 & 0.1253 & 0.2556 & 0.765 \\
MiniMind2     & 0.1282 & 0.1670 & 0.1219 & --    \\
\bottomrule
\end{tabular}
\end{table}

In the EBK task, GPT-4o leads across metrics, especially in METEOR (0.4094) and MMS (0.832), highlighting its strength in generating accurate and semantically coherent ECG-related knowledge. While DeepSeek-V3 and Qwen2.5 show moderate gains, they still trail in lexical and semantic precision. Notably, MiniMind2 achieves the highest BLEU-1 score (0.2842), indicating strong word-level overlap, but its lower METEOR and missing MMS suggest limited semantic richness. This positions MiniMind2 as effective for keyword-focused tasks, whereas GPT-4o remains superior for clinically reliable generation.

% ----- Table for Task GRA -----
\begin{table}[h]
\centering
\caption{Evaluation Metrics for Task GRA}
\label{tab:metrics_gra}
\begin{tabular}{lcccc}
\toprule
Model & BLEU-1 & ROUGE-L & METEOR & MMS \\
\midrule
GPT-4o        & 0.0139 & 0.1382 & 0.0781 & 0.693 \\
DeepSeek-V3   & 0.0247 & 0.1439 & 0.0680 & \cellcolor[rgb]{0.7,1,0.7}0.807 \\
Qwen2.5       & \cellcolor[rgb]{0.7,1,0.7}0.1429 & \cellcolor[rgb]{0.7,1,0.7}0.1883 & \cellcolor[rgb]{0.7,1,0.7}0.1730 & 0.766 \\
\bottomrule
\end{tabular}
\end{table}

In the GRA task, DeepSeek-V3 leads with MMS (0.807), indicating its superior ability to generate accurate risk assessments. GPT-4o follows with an MMS score of 0.693, performing well in ROUGE-L but struggling to fully capture risk-related nuances. Qwen2.5 in MMS (0.766), outperforming GPT-4o but still trailing DeepSeek-V3. Overall, DeepSeek-V3 excels in risk assessment generation, while GPT-4o and Qwen2.5 are stronger in lexical precision.

% ----- Table for Task LTD -----
\begin{table}[h]
\centering
\caption{Evaluation Metrics for Task LTD}
\label{tab:metrics_ltd}
\begin{tabular}{lcccc}
\toprule
Model        & BLEU-1          & ROUGE-L         & METEOR          & MMS    \\
\midrule
GPT-4o       & 0.1860          & \cellcolor[rgb]{0.7,1,0.7}0.2733 & 0.1994          & 0.793  \\
DeepSeek-V3  & 0.1895          & 0.2621          & 0.1780          & 0.741  \\
Qwen2.5      & \cellcolor[rgb]{0.7,1,0.7}0.2264 & 0.2379          & \cellcolor[rgb]{0.7,1,0.7}0.2870 & \cellcolor[rgb]{0.7,1,0.7}0.764 \\
\bottomrule
\end{tabular}
\end{table}

In the LTD task, Qwen2.5 stands out with an MMS score of 0.764 and a METEOR score of 0.2870, demonstrating its ability to handle longer text more effectively. GPT-4o and DeepSeek-V3 have lower scores, indicating challenges in generating coherent and contextually accurate long-form medical texts. Qwen2.5 seems to have an advantage in understanding and generating longer sequences, providing more coherent responses.

% ----- Table for Task PP -----
\begin{table}[h]
\centering
\caption{Evaluation Metrics for Task PP}
\label{tab:metrics_pp}
\begin{tabular}{lcccc}
\toprule
Model        & BLEU-1          & ROUGE-L         & METEOR          & MMS    \\
\midrule
GPT-4o       & \cellcolor[rgb]{0.7,1,0.7}0.2008 & \cellcolor[rgb]{0.7,1,0.7}0.3011 & \cellcolor[rgb]{0.7,1,0.7}0.3427 & \cellcolor[rgb]{0.7,1,0.7}0.816 \\
DeepSeek-V3  & 0.0772          & 0.1779          & 0.1815          & 0.783  \\
Qwen2.5      & 0.0491          & 0.0954          & 0.1957          & 0.764  \\
MiniMind2    & 0.1230          & 0.1336          & 0.1252          & --     \\
\bottomrule
\end{tabular}
\end{table}

In the PP (Patient Prognosis) task, GPT-4o outperforms all models across the board, with top scores in METEOR (0.3427) and MMS (0.816), reflecting its strong capability in generating fluent and contextually appropriate prognostic statements. DeepSeek-V3 and Qwen2.5 lag significantly, especially in BLEU-1 and ROUGE-L, suggesting less accurate predictions. MiniMind2 shows moderate lexical performance (BLEU-1: 0.1230), but its low METEOR (0.1252) and lack of MMS indicate weaker semantic depth. Overall, GPT-4o remains the most reliable for prognosis generation, while MiniMind2 may suit simpler summarization tasks.

% ----- Table for Task PRTK -----
\begin{table}[h]
\centering
\caption{Evaluation Metrics for Task PRTK}
\label{tab:metrics_prtk}
\begin{tabular}{lcccc}
\toprule
Model        & BLEU-1          & ROUGE-L         & METEOR          & MMS    \\
\midrule
GPT-4o       & \cellcolor[rgb]{0.7,1,0.7}0.3855 & \cellcolor[rgb]{0.7,1,0.7}0.4512 & \cellcolor[rgb]{0.7,1,0.7}0.4730 & \cellcolor[rgb]{0.7,1,0.7}0.781 \\
DeepSeek-V3  & 0.1562          & 0.2387          & 0.3550          & 0.760  \\
Qwen2.5      & 0.1276          & 0.1919          & 0.3107          & 0.756  \\
\bottomrule
\end{tabular}
\end{table}

In the PRTK task, GPT-4o dominates with the highest scores in MMS (0.781) and METEOR (0.4730), showing strong ethical communication abilities and an understanding of patient rights. DeepSeek-V3 and Qwen2.5 have relatively lower scores, especially in MMS, reflecting their weaknesses in generating responses that align with ethical guidelines and provide accurate prognostic information.

\subsection*{MiniMind2 Performance Comparison Across Metrics}

\begin{figure}[h]
    \centering
    \includegraphics[width=0.82\linewidth]{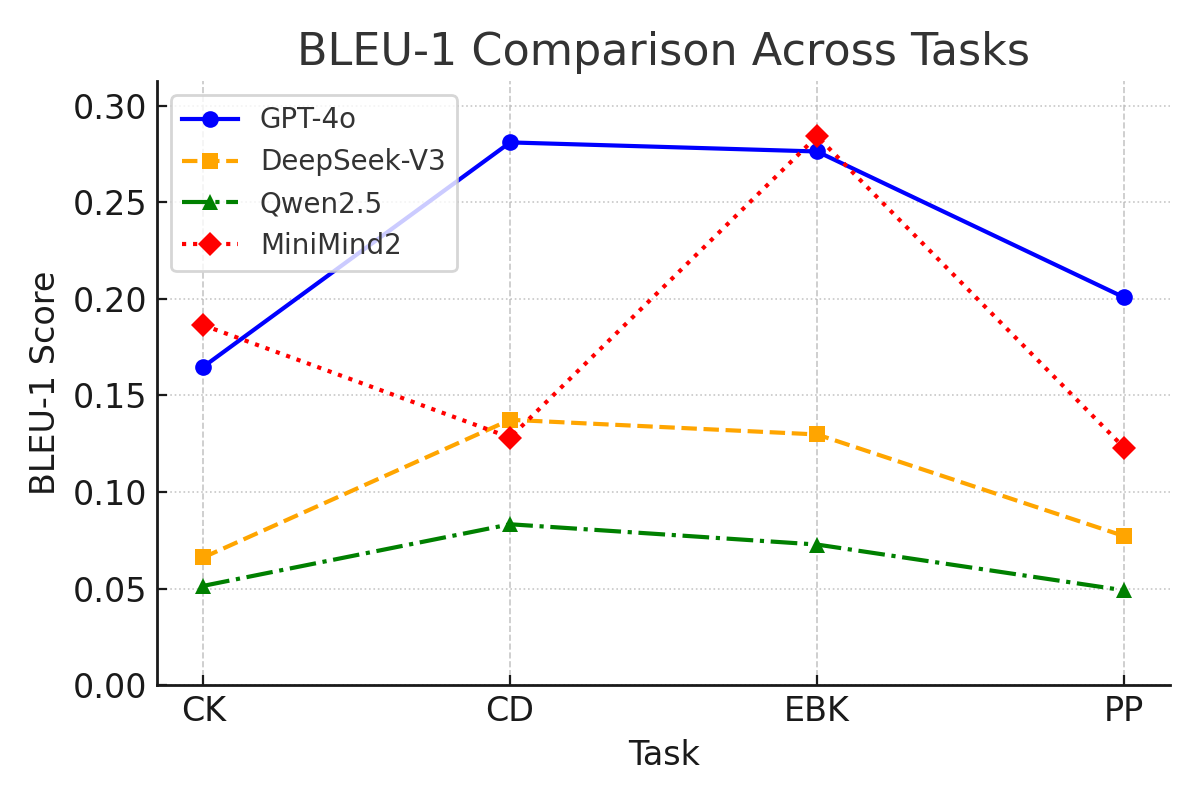}
    \includegraphics[width=0.82\linewidth]{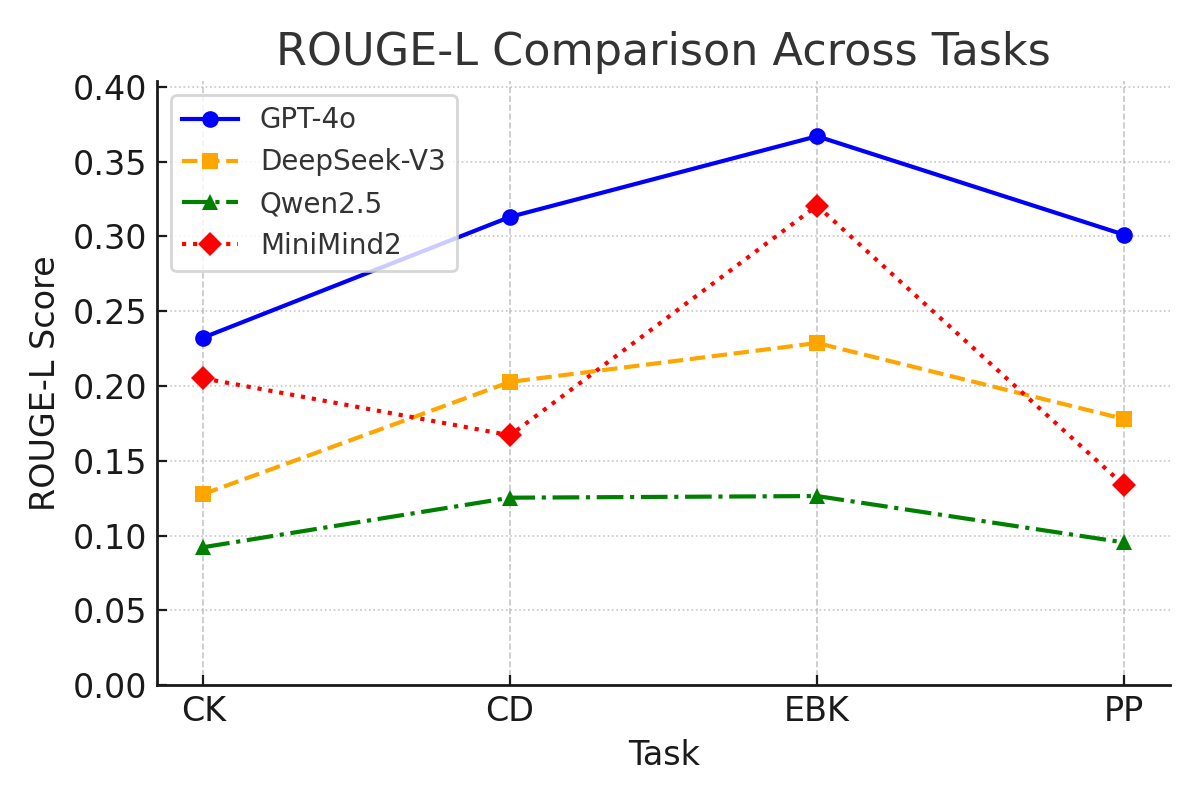}
    \includegraphics[width=0.82\linewidth]{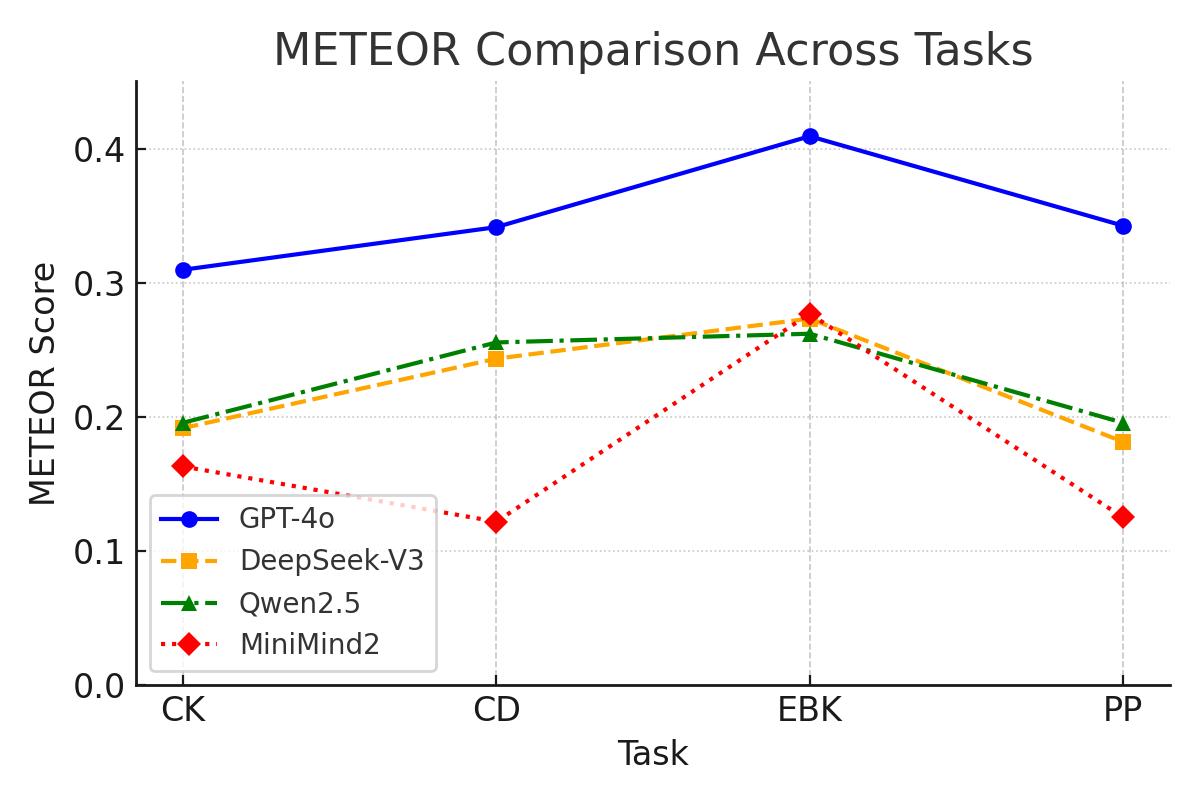}
    \caption{Performance comparison of MiniMind2 with larger models (GPT-4o, DeepSeek-V3, Qwen2.5) across BLEU-1, ROUGE-L, and METEOR on selected tasks}
    \label{fig10}
\end{figure}

In the evaluation of CK, CD, EBK, and PP, the lightweight MiniMind2 model (25.8M parameters) demonstrated notable capabilities despite its small size. After fine-tuning on ECG-Expert-QA, MiniMind2 achieved performance in knowledge-based tasks (CK and EBK) that was nearly comparable to much larger models. In CK, it even achieved a higher BLEU-1 score (0.1864) than GPT-4o (0.1648), indicating strong lexical alignment. However, its METEOR score was only 0.1631, less than half of GPT-4o's 0.3098, revealing limitations in semantic completeness. A similar pattern was observed in EBK, where MiniMind2's performance approached that of GPT-4o and DeepSeek-V3.

In contrast, for CD and PP, MiniMind2 underperformed, with METEOR scores around 0.12, while GPT-4o reached approximately 0.34. This suggests that MiniMind2 struggles with tasks requiring complex reasoning and richer language generation.

Nevertheless, given its compact size, MiniMind2 still delivered competitive results. In some lexical metrics like BLEU-1, it outperformed Qwen2.5 (e.g., in CK and EBK), and approached DeepSeek-V3 in basic QA tasks. These results indicate that high-quality, domain-specific datasets like ECG-Expert-QA can substantially boost the effectiveness of small models in targeted medical tasks, though significant gaps remain in complex diagnostic reasoning and semantic richness.

\subsection*{Summary} 
Among the four models (GPT-4o, DeepSeek-V3, Qwen2.5, and MiniMind2) evaluated on seven tasks from the ECG-Expert-QA benchmark, GPT-4o achieves the highest overall performance. It consistently attains top scores on both lexical metrics (e.g., BLEU-1) and semantic measures (e.g., METEOR and MMS), indicating a superior grasp of clinical context. DeepSeek-V3 and Qwen2.5 show moderate performance, each excelling in certain areas: DeepSeek-V3 is notably strong in risk assessment tasks, whereas Qwen2.5 handles long-form diagnostic narratives more effectively. Notably, despite its significantly smaller size (25.8M parameters), the fine-tuned MiniMind2 model delivers surprisingly high lexical accuracy, particularly on knowledge-based tasks like CK and EBK. However, MiniMind2 exhibits clear limitations in complex reasoning and in producing semantically rich responses compared to the larger models.

\section{Conclusion} 
\label{sec6} 
In this work, we introduce ECG-Expert-QA, a comprehensive benchmark designed to assess the performance of large language models (LLMs) in intelligent ECG interpretation. By combining real clinical data with synthetically generated cases, the dataset covers 12 diagnostic tasks and supports multi-turn, knowledge-intensive dialogues. It provides a solid foundation for evaluating both clinical reasoning and conversational capabilities in medical AI.

To assess the effectiveness of current LLMs in ECG-related diagnostic tasks, we developed a structured evaluation framework based on four widely recognized natural language generation (NLG) metrics: BLEU-1, ROUGE-L, METEOR, and the newly introduced MMS. Experiments with three representative models—GPT-4o, DeepSeek-V3, and Qwen2.5—revealed significant variations in performance across tasks and evaluation metrics. GPT-4o consistently delivered superior results, especially in complex, semantically rich, and ethically sensitive scenarios. Qwen2.5 showed particular strengths in handling long-form text and risk assessment tasks. DeepSeek-V3 maintained stable performance in knowledge-driven tasks but lagged in terms of semantic reasoning and adaptability to complex scenarios.

Our findings underscore the importance of multi-dimensional evaluation in medical AI, as different models excel in specific areas depending on task complexity, contextual depth, and ethical sensitivity. Furthermore, ECG-Expert-QA and its evaluation framework provide a crucial step towards the development of more reliable, explainable, and patient-centered AI systems for healthcare.

Looking ahead, future research should focus on integrating temporal ECG dynamics, real-time clinical workflows, and cross-lingual adaptability. Additionally, there is a need to enhance continual learning capabilities and safety assessments in LLM-based diagnostic tools. As an open-source resource, ECG-Expert-QA is poised to drive advancements in multimodal, conversational, and ethically conscious AI for healthcare.

\section{Future Work}

While ECG-Expert-QA provides a comprehensive benchmark for evaluating diagnostic reasoning in ECG-focused medical LLMs, several directions remain open for future exploration. First, expanding the benchmark to cover other cardiac-related modalities (e.g., phonocardiogram, echocardiography) and systemic conditions (e.g., metabolic or respiratory diseases) may enhance the versatility of evaluation scenarios.

Second, incorporating real-time and continuous monitoring data into the QA framework could further bridge the gap between static diagnosis and dynamic clinical decision-making. Future research could also explore the integration of wearable ECG devices for continuous multimodal monitoring, enabling more dynamic clinical modeling beyond static diagnostic tasks~\cite{bibTV}. Such an extension would support patient-specific longitudinal assessment, particularly for chronic conditions like heart failure, arrhythmia recurrence, or cardiopulmonary interactions.

Lastly, the alignment between expert feedback and model-generated QA paths remains a key frontier. We envision fine-tuning LLMs using reinforcement learning with expert trajectories, and further enhancing trustworthiness through explainability and counterfactual simulations embedded in multi-agent interactions.

\end{document}